\documentclass[twocolumn, secnumarabic, amssymb, amsmath,nobibnotes, aps, prl,nofootinbib]{revtex4}
\begin{document}
\title{The Thermodynamics Governing \lq Endoreversible\rq\ Engines}
\author{B. H. Lavenda}
\email{bernard.lavenda@unicam.it}
\affiliation{Universit$\grave{a}$ degli Studi, Camerino 62032 (MC) Italy}
\date{\today}
\newcommand{\sumn}{\sum_{i=1}^{n}\,}
\newcommand{\sumk}{\sum_{i=1}^{k}\,}
\newcommand{\half}{\mbox{\small{$\frac{1}{2}$}}}
\newcommand{\fourth}{\mbox{\small{$\frac{1}{4}$}}}\newcommand{\twothirds}{\mbox{\small{$\frac{2}{3}$}}}
\newcommand{\nn}{\mbox{\small{$\frac{1}{n}$}}}
\begin{abstract}
The thermodynamics of the Curzon-Ahlborn engine, which is a prototype of endoreversible engines, is elucidated. In particular, their criterion for adiabatic equilibration is revised. The so-called irreversibility of endoreversible engines arises from the selection of the coldest reservoir for heat rejection. Rather, if the reservoirs are allowed to come into thermal and mechanical contact, a mean value results which optimizes the work output and heat uptake, and is entirely reversible. The Carnot efficiency cannot be beaten because nothing is as cold as the coldest reservoir.
\end{abstract}
 
\maketitle
\section{The criterion of reversibility in an endoreversible engine}
An endoreversible engine has been defined as an \lq\lq irreversible engine where all irreversibilities are restricted to the coupling of the engine to the external world\rq\rq \cite{Rubin}. The Curzon-Ahlborn \cite{CA} engine is supposedly \cite{deVos} a prototype of an endoreversible engine.\par
The Curzon-Ahlborn engine makes the distinction between the temperatures of the reservoirs and the temperatures of the working fluid at which heat is absorbed and rejected. Whereas the temperature of the hot reservoir is $T_1$, a quantity of heat $Q_1$ is absorbed at the fluid temperature, $T_{1w}<T_1$. The isothermal expansion step in Carnot's cycle is replaced by a discrete form of Fourier's law
\[
Q_1=g_1\left(T_1-T_{1w}\right), \]
where the coefficient of proportionality, $g_1$, is the thermal conductance. Likewise, a quantity of heat, $Q_2>0$, is given up to the cold reservoir which is at temperature, $T_2$. However, the fluid is at the higher temperature $T_{2w}$. The isothermal compression step in Carnot's cycle is replaced by
\[
Q_2=g_2\left(T_{2w}-T_2\right), \]
where $g_2$ is another conductance.\par
The crucial step in the Curzon-Ahlborn argument is their condition of reversibility:
\begin{equation}
\frac{Q_1}{T_{1w}}=\frac{Q_2}{T_{2w}}, \label{eq:CA}
\end{equation}
which is taken as the condition for entropy conservation \cite[p. 34]{deVos}. However, (\ref{eq:CA}) cannot be the condition for entropy conservation because it is {\bf{not\/}} the sum of ratios of the heats to the temperatures which they are at that is to vanish in a reversible cycle (in this formula heats are written as algebraic quantities), i.e.,
\begin{equation}
\sumn \frac{Q_i}{T_i}=0. \label{eq:II}
\end{equation}
Certainly, (\ref{eq:II}) is not (\ref{eq:CA}) (with the heat rejected positive in (\ref{eq:CA})). Even a non-vanishing total entropy change is calculated for the Curzon-Ahlborn engine \cite[p. 36]{deVos}. So it is a fair question to ask what is (\ref{eq:CA})?\par
According to (\ref{eq:CA}), the efficiency depends upon the ratio of the intermediate temperatures and is something to vary in order to maximize the work output. By varying the work with respect to the efficiency, $\eta=1-T_{2w}/T_{1w}$, Curzon and Ahlborn obtain an optimal efficiency,
\begin{equation}
\eta=1-\sqrt{\frac{T_2}{T_1}}, \label{eq:eff-CA}
\end{equation}
which clearly inferior to the Carnot efficiency, $\eta_C=1-T_2/T_1$. With this optimal efficiency, they obtain maximum work output
\begin{equation}
W=g\left(\sqrt{T_1}-\sqrt{T_2}\right)^2, \label{eq:W-CA}
\end{equation}
where $g$ is half of the harmonic mean of the conductances. Since $W=\eta Q_1$, these authors find the maximum heat intake as
\begin{equation}
Q=g\sqrt{T_1}\left(\sqrt{T_1}-\sqrt{T_2}\right). \label{eq:Q-CA}
\end{equation}
\par
The conceptual error is (\ref{eq:CA}). What (\ref{eq:CA}) should say is this: An engine absorbs a quantity of heat, $Q_1$, at temperature $T_1$, and emits the quantity of heat $\left(\mathfrak{M}(T)/T_1\right)Q_1$ at temperature $\mathfrak{M}(T)$, viz.,
\begin{equation}
Q_2=\frac{\mathfrak{M}(T)}{T_1}Q_1. \label{eq:CA-bis}
\end{equation}
What Curzon and Ahlborn rediscovered was that the final mean temperature at which adiabatic equilibration occurs yields maximum work. This is implicit in the 1967 paper by Cashwell and Everett \cite{CE} in which they showed that the final mean temperature determined from the first law is greater than that determined by the second law. Therefore, the second law follows from the property that means are monotonically increasing functions of their order.\par
Here we invert their argument: We use adiabatic equilibration of the second law and derive maximum work from the first law. The Curzon-Ahlborn analysis has nothing to do with maximizing power at finite rates. The introduction of time is misleading. There is no time frame in which the final temperature to which the system equilibrates adiabatically, $\mathfrak{M}(T)$. We shall show that this common temperature renders the work a maximum. The mean temperature is always greater than the final temperature of the coldest reservoir. It is this fact that leads to an illusory increase in entropy.  (\ref{eq:CA-bis}) is a statement of reversibility according to the second law (taking into account that we take the heat rejected to the colder reservoir as positive in contrast to (\ref{eq:II})). The optimality of the Carnot efficiency is simply a consequence of the fact that the temperature of adiabatic equilibration is greater than the temperature of the coldest reservoir: \lq nothing can be colder than the coldest reservoir.\rq\par
\section{Thermodynamics of means}
In order to proceed with the development, let us recall some of the salient features of the thermodynamics of means \cite{CE,Lav}.
According to the \lq doctrine\rq\ \cite{Truesdell} of latent, $M_i$, and specific, $N_i$, heats,  the quantity of heat, $dQ$, needed to alter the volume, $V$, to $V+dV$, and absolute temperature $T$ to $T+dT$, of a system composed of $n$ components is \cite{Kelvin}
\begin{equation}
dQ=\sumn \left(M_i\,dV+N_i\,dT\right). \label{eq:dQ}
\end{equation}
The first law expresses the fact that the internal energy, $dE=dQ-p\,dV$ is a total differential, resulting in
\begin{equation}
\frac{\partial p}{\partial T}=\sumn\left(\frac{\partial M_i}{\partial T}-
\frac{\partial N_i}{\partial V}\right), \label{eq:I}
\end{equation}
while the second laws identify $V^s$ and $1/T$ as integrating factors for (\ref{eq:dQ}). The second laws can thus be stated as
\begin{equation}
\frac{\partial p}{\partial T}=\frac{\sumn M_i}{T}=\frac{s\sumn N_i}{V}. \label{eq:II-L}
\end{equation}
The second equality identifies $s$ as the Gr\"uneisen parameter, which varies from $\mbox{\small$\frac{2}{3}$}$ for a nonrelativistic gas to $\mbox{\small$\frac{1}{3}$}$ for an ultrarelativistic one.\par
The first equality in (\ref{eq:II-L}) had a much more prominent role in the history of thermodynamics, being the Carnot equation and $\mu=1/T$, the Carnot function. It was Thomson who determined $\mu$, \begin{quote} a quantity which has an absolute value, the same for all substances for any given temperature, but which may vary with temperature in a manner that can only be determined by experiment.
\end{quote} Thomson credits Carnot with the remarkable theorem that the ratio of the work done to the heat absorbed in an isothermal transition, $(\partial p/\partial T)\,dT dV/\sumn M_i\,dV$ is the same for all substances at the same temperature. An analogous statement can be said about the second inequality in (\ref{eq:II-L}) \cite{Lav}: \begin{quote} The ratio of the work done to the heat absorbed in an isochoric transition, $(\partial p/\partial T)\,dT dV/\sumn N_i\,dT$ is the same for all substances at the same volume.
\end{quote}\par
If the $n$-cells of masses $m_i$ and heat capacities $c_i(T)$, where $N_i=m_ic_i$, all initially at different temperatures $T_i$, are placed in thermal contact, the entire system being adiabatically isolated, then after an indefinite long period of time will reach a common mean temperature, $\mathfrak{M}$. This final temperature is determined from adiabatic equilibration:
\begin{equation}
\Delta S=\sumn\int_{T_i}^{\mathfrak{M}}\frac{N_i(t)}{t}\,dt=0. \label{eq:adeq}
\end{equation}
\par
The Curzon-Ahlborn result refers to a perfect gas where $N_i$ are constant, independent of both temperature and volume. Condition (\ref{eq:adeq}) then gives
\[\sumn N_i\ln\mathfrak{M}-N_i\ln T_i=0,\]
or, equivalently
\begin{equation}
\mathfrak{M}_0(T)=\left(\prod_{i=1}^n T_i^{N_i}\right)^{1/\sumn N_i}, \label{eq:geometric}
\end{equation}
which identifies the final equilibrium temperature as the mean of order $0$, or the geometric mean.\par In the case of only two cells, or reservoirs, of identical masses and heat capacities, the final mean temperature will be $\mathfrak{M}_0(T)=\sqrt{T_1T_2}$, and introducing this into (\ref{eq:CA-bis}) gives
\[\frac{Q_2}{Q_1}=\sqrt{\frac{T_2}{T_1}},\]
the Curzon-Ahlborn result for the optimal efficiency.\par
Adiabatic equilibration (\ref{eq:adeq}), leading to the geometric mean as the lowest common temperature of the composite system, will result, according to the first law, in the work
\begin{equation}
W=\sumn\int_{\mathfrak{M}_0}^{T_i}N_i\,dt. \label{eq:W}
\end{equation}
For two cells, or reservoirs, having the same mass and heat capacity, $N_i=N$, (\ref{eq:W}) reduces to 
\[
W=2N\left(\mathfrak{M}_1-\mathfrak{M}_0\right)=N\left(\sqrt{T_1}-\sqrt{T_2}\right)^2, \]
corresponding to the Curzon-Ahlborn result, (\ref{eq:W-CA}). It could have been taken as a \lq thermodynamic proof\rq\ of the arithmetic-geometric mean inequality as far back as 1868 \cite{Tait}!
\par
According to Curzon-Ahlborn, we would be tempted to write the optimal relation, corresponding to (\ref{eq:CA}), as
\[\frac{Q_1}{\sqrt{T_1}}=\frac{Q_2}{\sqrt{T_2}}.\]
This would be in contradiction of the second law which, out of the entire set of admissible empirical temperatures, selects the absolute temperature. Were we to consider $T_1$ and $T_2$ as empirical temperatures, the resulting absolute temperature would increase as the square of the empirical temperature, contradicting the fact that the empirical temperature must increase at least as fast as the absolute temperature.\par
Hence, the correct criterion of reversibility is (\ref{eq:CA-bis}), in which $Q_2$  is the quantity of heat $Q_1$ at temperature $T_1$ that is emitted at temperature $\mathfrak{M}_0(T)$.  And since $\mathfrak{M}_0(T)>T_2$ it follows that
\[\frac{Q_1}{T_1}-\frac{Q_2}{T_2}<0,\]
giving rise to an apparent increase in entropy \cite[p. 36]{deVos}
\begin{equation}\Delta S=\left(\eta_C-\eta\right)\frac{Q_1}{T_2},\label{eq:S}
\end{equation}
where $\eta_C$ and $\eta$ are the Carnot and Curzon-Ahlborn, (\ref{eq:eff-CA}), efficiencies, respectively.
However, there is nothing irreversible about the Curzon-Ahlborn engine. It is only when we demand that the colder heat reservoir be at temperature $T_2$, and not at the mean temperature, $\sqrt{T_1T_2}$, does it appear to be irreversible, giving rise to a positive entropy change (\ref{eq:S}). 
\par
How close can we come to the Carnot efficiency? It is well-known that \cite{HLP}
\[\lim_{q\rightarrow-\infty}\mathfrak{M}_q(T)=\min T,\]
so that we might be tempted to modify the second law to read
\[\Delta S=\sumn\int_{T_i}^{\mathfrak{M}}\frac{N_i}{t^{\alpha}}\,dt\]
for some $\alpha>0$ in order to get a mean of negative order. However, if the second law is not to be violated $\alpha\le1$. From the fact that means are monotonically increasing functions of their order, an adiabatic equilibration yielding a mean temperature, $\mathfrak{M}_{1-\alpha}(T)$, would be greater than the geometric mean, $\mathfrak{M}_0(T)$, predicted by the second law.\par
\section{Isothermal \lq endoreversible\rq\ engine}
The Curzon-Ahlborn engine uses  isochoric processes to replace  isothermal ones. There is no reason to do so. For all other interactions than thermal, the first and second laws are incomparable \cite{Lav}. This is due to the fact that energy and (metrical) entropy are first-order homogeneous. What is requires is another adiabatic potential which is not a first-order homogeneous function for  this will allow us to compare means of different orders and thereby establish the maximum property.\par For the perfect gas, the metrical entropy, $S$, is related to the empirical entropy, $\sigma$, logarithmically,
\[S(\sigma)=R\ln\sigma^{1/s},\]
where $R$ is the gas  constant. The second law,
\[T\,dS(\sigma)=T\,S^{\prime}(\sigma)\,d\sigma=\frac{RT}{s\sigma}\,d\sigma=\frac{Rd\sigma}{sV^s}=dQ,\]
shows that $V^s$ is also an integrating factor for the heat \cite{Ein,Lav},
\[\frac{R}{s}d\sigma=V^s\,dQ=\sumn\,V^s\left(M_i\,dV+N_i\,dT\right),\] where $\sigma=TV^s$ are the adiabats. If the massses of all the cells are the same $M_i=p$, the adiabatic equilibration for an isothermal expansion, 
\begin{eqnarray*}
\Delta\sigma & = & sT\sumn\int_{V_i}^{\mathfrak{M}(V)}v^{s-1}\,dv\\
& = & T\left(n\mathfrak{M}^s(V)-\sumn V_i^s\right)=0,
\end{eqnarray*}
identifies the mean of order $s$,
\begin{equation}
\mathfrak{M}_s(V)=\left(\frac{1}{n}\sumn V_i^s\right)^{1/s},\label{eq:mean-s}
\end{equation}
as the final  volume of the composite system.\par
The maximum heat absorbed in this isothermal expansion is
\begin{equation}
Q=RT\sumn\int_{V_i}^{\mathfrak{M}_s}\,\frac{dv}{v}=nRT\,\ln\left(\frac{\mathfrak{M}_s}{
\mathfrak{M}_0}\right)>0, \label{eq:Q-iso}
\end{equation}
where 
\begin{equation}
\mathfrak{M}_0(V)=\left(\prod_{i=1}^n\,V_i\right)^{1/n}. \label{eq:geo-mean}
\end{equation}
 Inequality  (\ref{eq:Q-iso}) follows again from the facts that means are monotonically increasing functions of their order, and $s>0$. If the $n$ cells, initially at volumes $V_1,V_2,\ldots, V_n$ were brought into mechanical contact and left alone for an indefinite period of time, the final volume would be given by the geometrical mean (\ref{eq:geo-mean}). Taking a cue from probability theory, the volume would be proportional to the probability measure of a given set. The condition for statistical independence, sometimes referred to as a \lq mixing property\rq, would be the factorization of the measure into products of individual measures, whose mean value would be given by (\ref{eq:geo-mean}). This would be the smallest volume conceivable.\par
\section{Maximum work output and heat uptake}
In this section we establish that the final mean temperature and volume, determined from adiabatic equilibration, yields maximum work output and maximum heat uptake with respect to all other admissible sets of temperatures and volumes.\par
Let us first consider the case of thermal interactions. Consider, again, a set of $n$ cells, or reservoirs, all at different temperatures, $T_i$ with $\max T_i=T_1$ and $\min T_i=T_2$. Following Cashwell and Everett \cite{CE}, we divide the reservoirs into two groups:  \lq$\ell$\rq\ for those with $T_i\le\mathfrak{M}$, and  \lq $u$\rq\ for those reservoirs having temperatures $T_i>\mathfrak{M}$. Denote by $\lambda(T)$ a continuous, monotonically \emph{decreasing\/} function of $T$. On the strength of the adiabaticity constraint, and the monotonicity of $\lambda(T)$, the following string of inequalities result
\begin{eqnarray*}
\lefteqn{\lambda(\mathfrak{M})\sum_{\ell}\int_{T_i}^{\mathfrak{M}}N_i(t)\,dt}\\
& \le & \sum_{\ell}\int_{T_i}^{\mathfrak{M}}\lambda(t) N_i(t)\,dt=\sum_{u}\int_{\mathfrak{M}}^{T_i}\lambda(t) N_i(t)\,dt\\
& \le & \lambda(\mathfrak{M})\sum_{u}\int_{\mathfrak{M}}^{T_i}N_i(t)\,dt.
\end{eqnarray*}
In other words, the adiabatic equilibration condition (\ref{eq:CA-bis}) could have equally as well been written as $Q_1=\mathfrak{M}(T)Q_2/T_2$.
From the string of inequalities, we therefore conclude
\begin{equation}
\sumn\int_{\mathfrak{M}}^{T_i}N_i(t)\,dt\ge0, \label{eq:ineq-T1}
\end{equation}
with equality  holding if $T_1=T_2$.\par
Let $\hat{T}_i$ denote any admissible set of final temperatures of the reservoirs. Since
\[\sumn\int_{T_i}^{\hat{T}_i}N_i(t)\,dt=\sumn\int_{T_i}^{\mathfrak{M}}N_i(t)\,dt+
\sumn\int_{\mathfrak{M}}^{\hat{T}_i}N_i(t)\,dt,\]
and the fact that (\ref{eq:ineq-T1}) guarantees that the last term will be positive, regardless of the nature of the set of temperatures, it follows that
\begin{equation}
\sumn\int_{\mathfrak{M}}^{T_i}N_i(t)\,dt>\sumn\int_{\hat{T}_i}^{T_i}N_i(t)\,dt.
\label{eq:ineq-T2}
\end{equation}
Consequently, the work output will be maximum; this establishes the validity of the revised Curzon-Ahlborn condition (\ref{eq:CA-bis}).\par
The principle of maximum heat uptake in an isothermal process can be established in a similar manner. Let $\kappa(V)$ be a continuous, monotonically \emph{increasing\/} function of the volume. Out of the set of positive numbers, $V_1,V_2,\ldots, V_n$, suppose $\max V_i=V_2$ and $\min V_i=V_1$. Again dividing the reservoirs into two group depending upon whether the $V_i$ are less or greater than the mean value, $\mathfrak{M}$, we get the reverse string of inequalities
\begin{eqnarray*}
\lefteqn{\kappa(\mathfrak{M})\sum_{\ell}\int_{V_i}^{\mathfrak{M}}M_i(v)\,dv}\\
&\ge & \sum_{\ell}\int_{V_i}^{\mathfrak{M}}\kappa(v)M_i(v)\,dv=\sum_{u}\int_{\mathfrak{M}}^{V_i}\kappa(v)M_i(v)\,dv\\
&\ge &\kappa(\mathfrak{M})\sum_{u}\int_{\mathfrak{M}}^{V_i}M_i(v)\,dv,
\end{eqnarray*}
from which it follows
\begin{equation}
\sumn\int_{V_i}^{\mathfrak{M}}M_i(v)\,dv\ge0, \label{eq:ineq-V1}
\end{equation}
where the equality holds if $V_1=V_2$. If we consider any other set of final volumes, $\hat{V}_i$, then
\begin{eqnarray*}
\lefteqn{\sumn\int_{V_i}^{\hat{V}_i}M_i(v)\,dv}\\
& = &
\sumn\int_{V_i}^{\mathfrak{M}}M_i(v)\,dv-\sumn\int_{\hat{V}_i}^{\mathfrak{M}}M_i(v)\,dv.
\end{eqnarray*}
Since inequality (\ref{eq:ineq-V1}) ensures that the last term is positive, whatever be the initial states, we conclude that
\[
\sumn\int_{V_i}^{\mathfrak{M}}M_i(v)\,dv\ge\sumn\int_{V_i}^{\hat{V}_i}M_i(v)\,dv.
\]
This establishes the maximum property of the heat uptake in an isothermal process when the final state has the volume given by (\ref{eq:mean-s}).\par
\section{Comparison of \lq endoreversible\rq\ engines}
Comparison may now be made between the isochoric engine of Curzon-Ahlborn and the isothermal engine of the last section. The efficiency of the engine is
\begin{equation}
\eta=1-\frac{\mathfrak{M}^s_0(V)}{\mathfrak{M}^s_s(V)}. \label{eq:eta-iso}
\end{equation}
The maximum work output is the product of (\ref{eq:eta-iso}) and (\ref{eq:Q-iso}), viz.,
\begin{equation}
W=\eta Q_1=\frac{n}{s}RT_1\left(1-\frac{\mathfrak{M}_0^s(V)}{\mathfrak{M}_s^s(V)}\right)\ln\left(\frac{\mathfrak{M}_s^s(V)}{\mathfrak{M}_0^s(V)}\right). \label{eq:W-iso}
\end{equation}
\par
In order to make a comparison of the two engines, we set $n=2$, and use the adiabatic constraints $V_1^sT_1=\mbox{const.}$, and $V_2^sT_2=\mbox{const.}$ These convert the mean of order $s$, (\ref{eq:mean-s}), for the volume into the inverse of the harmonic mean for the temperature
\begin{eqnarray}
\lefteqn{\mathfrak{M}_s^s(V)=\half\left(V_1^s+V_2^s\right)} \nonumber\\ & = & \half\left(\frac{1}{T_1}+\frac{1}{T_2}\right)= \frac{\mathfrak{M}_1(T)}{\mathfrak{M}_0^2(T)},\label{eq:M1}
\end{eqnarray}
and the geometric mean of the volume, (\ref{eq:geo-mean}), into the inverse of geometric mean of the temperature,
\begin{equation}\mathfrak{M}_0^s(V)=\sqrt{V_1^sV_2^s}=\frac{1}{\sqrt{T_1T_2}}=1/\mathfrak{M}_0(T).\label{eq:M2}
\end{equation}
Since we will be concerned only with their ratio, we have dispensed with the arbitrary constants in (\ref{eq:M1}) and (\ref{eq:M2}).
\par
Expressed in terms of the temperature, the efficiency (\ref{eq:eta-iso}) is
\[
\eta=1-\frac{\mathfrak{M}_0(T)}{\mathfrak{M}_1(T)}\]
 The maximum heat uptake,
\[Q_1=\frac{2}{s}R\mathfrak{M}_{\infty}(T)\ln\left(\frac{\mathfrak{M}_1(T)}{\mathfrak{M}_0(T)}\right),\]
which for small differences between arithmetic and geometric means can be approximated by
\[
Q_1\simeq\frac{2}{s}R\mathfrak{M}_{\infty}(T)\frac{\left(\mathfrak{M}_1(T)-\mathfrak{M}_0(T)\right)}{\mathfrak{M}_0(T)},  \]
where $\mathfrak{M}_{\infty}(T)=T_1$.
Therefore, the approximate expression for the maximum work is
\[
W\simeq\frac{2}{s}R\mathfrak{M}_{\infty}(T)\frac{\left(\mathfrak{M}_1(T)-\mathfrak{M}_0(T)\right)^2}
{\mathfrak{M}_0(T)\mathfrak{M}_1(T)}. \]
\par
For comparison purposes we express the efficiency (\ref{eq:eff-CA}), heat uptake, (\ref{eq:Q-CA}), and work output, (\ref{eq:W-CA}), of the Curzon-Ahlborn engine in terms of mean values,
\begin{equation}
\eta=1-\frac{\mathfrak{M}_0(T)}{\mathfrak{M}_{\infty}(T)}, \label{eq:eta-CA}
\end{equation}
\[
Q=g\mathfrak{M}_{\infty}(T)\frac{\mathfrak{M}_1(T)-\mathfrak{M}_0(T)}{T_1-\mathfrak{M}_0(T)}, \]
and
\[
W=g\left(\mathfrak{M}_1(T)-\mathfrak{M}_0(T)\right), \]
respectively.
 For the West Thurrock Coal Fired Steam Plant, which has an efficiency of $36\%$, the Carnot efficiency is $64\%$, while the Curzon-Ahlborn and isothermal efficiencies are $40\%$ and $12\%$, respectively. In other words, in this particular example, the efficiency of the isothermal engine is only $30\%$ as efficient as the isochoric one.\par
Moreover, if we set the Curzon-Ahlorn conductance $g$ equal to the total specific heat, $g=R/s$, then the heat uptakes of the two engines are roughly the same, for the same power plant under discussion, while the work output of the isothermal engine is only $40\%$ that of the Curzon-Ahlborn engine.\par
\section{Conclusions}
Rubin's definition of the Curzon-Ahlborn engine is wanting. The irreversibilities which supposedly arise from the \lq coupling of the engine to the external world\rq\ are fictitious. Maximum work output and heat uptake arise when the reservoirs, initially at different temperatures or volumes, are allowed to interact, thermally or mechanically, and whose final mean values are determined by adiabatic equilibration. These mean values are necessarily greater than the coldest reservoir or smallest volume so that the efficiencies are less than the Carnot efficiency. The Carnot efficiency singles out these extreme reservoirs, and nothing can be colder, or smaller, than these extremes. The illusory increase in entropy is due to the fact that the heat rejected to the cold reservoir is divided by the temperature of that reservoir, which is definitely lower than the mean value obtained from adiabatic equilibration.

\end{document}